# Machine learning for laser-induced electron diffraction imaging of molecular structures

Xinyao Liu[1], Kasra Amini[1], Aurelien Sanchez[1], Blanca Belsa[1], Tobias Steinle[1], Jens Biegert[1,2,†]

[1]ICFO - Institut de Ciencies Fotoniques, The Barcelona Institute of Science and Technology, 08860 Castelldefels (Barcelona), Spain.
[2]ICREA, Pg. Lluís Companys 23, 08010 Barcelona, Spain.

[†]To whom correspondence should be addressed to. Email: jens.biegert@icfo.eu

**Abstract**

Ultrafast diffraction imaging is a powerful tool to retrieve the geometric structure of gas-phase molecules with combined picometre spatial and attosecond temporal resolution. However, structural retrieval becomes progressively difficult with increasing structural complexity, given that a global extremum must be found in a multi-dimensional solution space. Worse, pre-calculating many thousands of molecular configurations for all orientations becomes simply intractable. As a remedy, here, we propose a machine learning algorithm with a convolutional neural network which can be trained with a limited set of molecular configurations. We demonstrate structural retrieval of a complex and large molecule, Fenchone ($C_{10}H_{16}O$), from laser-induced electron diffraction (LIED) data without fitting algorithms or *ab initio* calculations. Retrieval of such a large molecular structure is not possible with other variants of LIED or ultrafast electron diffraction. Combining electron diffraction with machine learning presents new opportunities to image complex and larger molecules in static and time-resolved studies.

**Introduction**

The retrieval of complex molecular structures with electron or X-ray diffraction is challenging due to the multi-dimensional solution space in which a global extremum has to be found[1] to extract structural information from the diffraction data. Simple convergence strategies are readily implemented for small molecules, but such methods quickly become intractable for complex systems. For example, relativistic 3.7 MeV ultrafast electron diffraction (UED) time-resolved the ring-opening reaction of 1,3-cyclohexadiene, but the method could not identify the complex transient structure beyond 3 Å, which is expected to arise from three possible isomers[2]. X-ray diffraction imaging with free-electron laser pulses requires a careful balance between sufficient beam brightness and avoiding structural damage by the strong X-ray pulses[3–5]. Such issues severely limit the number of studies and necessitate *ab initio* calculations[6–8]. Laser-induced electron diffraction (LIED)[1,9–22] is a powerful laser-based UED method that images even singular molecular structures with combined sub-atomic picometre and femtosecond-to-attosecond spatiotemporal resolution. Here the challenge of structural retrieval arises from the strong-field nature of self-imaging the structure[1,16–18,20,23,24] by recolliding a laser-driven attosecond wavepacket after photo ionization. In LIED, an electron wave packet is: (i) tunnel ionized from the parent molecule in the presence of strong laser field; (ii) accelerated and driven back by the oscillating electric field of the laser; and (iii) rescattered against the parent ion's atomic cores. The geometrical information of the target nuclei is encoded in the detected momentum distribution of rescattered electrons. LIED often uses the quantitative rescattering (QRS) theory[13,14] to retrieve the structure of simple molecular systems. The QRS theory enables the extraction of field-free elastic electron scattering cross-sections from electron rescattering measurements performed under the presence of a strong laser field. The molecular structure is embedded onto the momentum distribution of the highly-energetic rescattering electrons. But for large and complex molecular structures, the QRS retrieval quickly becomes intractable due to the difficulty in identifying a unique solution in the multi-dimensional solution space. A somewhat remedy is found by reducing the dimensionality of the problem with FT-LIED[1,17,24,25]. Nevertheless, a multi-peak fitting procedure is needed to identify bond distances. Such an approach becomes ambiguous when the radial distribution function does not exhibit clear and separable structures. We note that similar problems arise in essentially all implementations of structural imaging and many retrieval algorithms severely limit the size of the molecular system under investigation.

To overcome these limitations, we employ a machine learning (ML) algorithm for LIED (ML-LIED) to accurately extract the three-dimensional (3D) molecular structure of larger and more complex molecules. The ML-LIED method avoids the use of chi-square fitting algorithms, multi-peak identification procedures, and *ab initio* calculations. The method draws from the interpolation and learning capabilities of ML which significantly reduces the required molecular configurations to train the system for a much larger solution space. We demonstrate our method's capability by extracting a single accurate molecular structure with picometre sub-atomic spatial resolution on the symmetric top linear system acetylene ($C_2H_2$), an asymmetric top 2D system, carbon disulfide ($CS_2$) and a complex large 3D system, (+)-Fenchone ($C_{10}H_{16}O$). For the specific problem of LIED, we employ ML with a convolutional neural network (CNN). Comparing with the commonly used artificial neural networks such as a fully connected neural network[26] or recurrent neural network (RNN)[27], the CNN is well suited for problems in image recognition to identify subtle features from an image at different levels of complexity similar to a human brain[28,29]. Machine learning algorithms are usually trained either with supervised or unsupervised learning. Under the supervised learning scenario, both the classification and regression methods are commonly chosen. The classification method is generally used to predict a specific category of data, such as in facial recognition[29] and crystal determination[30]. Here, we implement the regression method to quantitatively extract the structural parameters that identify the molecular structure. The molecular structure is found via the relationship between the molecular configuration and the molecular interference signal through the corresponding two-dimensional differential cross sections (2D-DCSs) from its database. Using 2D-DCSs as an input, our ML algorithm takes full advantage of the complete molecular interference signal rather than only considering a one-dimensional portion of the interference signal as is typically used in other methods[16,17]. Moreover, the ML CNN is capable to interpolate between samples in the database of pre-calculated structures to provide a meaningful configuration to measured data. This feature of ML CNN is especially important to identify large and complex molecular structures, since it is simply impossible to calculate all possible molecular structures with sufficient structural resolution and due to the many-fold degrees of freedom. We show a sufficiently reduced database suffices, which only considers (i) changes in a few important groups of atoms and (ii) a molecule-wide global change, allowing the algorithm to learn the relationship between the molecular structures and corresponding interference signals in our input database. Thus, our ML model, together with the CNN algorithm and regression method, provides a new way forward to identify the structure of large complex molecules and transient structures with similar geometric configurations in time-resolved pump-probe measurements.

This paper starts with a description of the machine learning scheme for LIED, followed by details on our machine learning model's training with a CNN and its subsequent validation. We then predict the molecular structure of $C_2H_2$ and $CS_2$ using the machine learning algorithm and experimental LIED data, followed by a comparison of the ML predicted molecular structures to those retrieved with the QRS method. We then demonstrate our ML model's capability to accurately predict the 3D molecular structure of (+)-Fenchone, which failed to retrieve with FT-LIED and QRS-LIED. Lastly, we discuss the advantages, limitations, and many-fold applications of the ML framework in retrieving large complex gas-phase molecular structures in the context of structural retrieval and time-resolved imaging of chemical reactions.

**Results**

**Machine learning scheme for LIED.** We employ machine learning (ML as an image recognition system to predict the measured molecular structure, a schematic of which is shown in Fig. 1. We start by generating a database containing the 3D cartesian coordinate of each atom in thousands of different molecular structures (as labels), spanning a coarse array of possible structures. For each structure, we generate the corresponding 2D-DCS map (stored as images in the database) by calculating the DCS of the elastic scattering of electrons on atoms in the molecule using the independent atomic model (IAM)[14,31,32]. The 2D-DCSs are calculated as a function of the electron's return energy (i.e., energy at the instance of rescattering) and rescattering angle (i.e. the change in angle caused by rescattering). The measured 2D-DCS signal, $\sigma_{tot}$, is comprised of two components: (i) the incoherent sum of atomic scatterings, $\sigma_{atom}$, and (ii) a modulating coherent molecular scattering signal, $\sigma_{\text{coherent}}$, which is approximately one order lower than $\sigma_{atom}$, i. e. $\sigma_{tot} = \sigma_{atom} + \sigma_{\text{coherent}}$. Here, the $\sigma_{atom}$ signal contributes as a background signal to our total scattering signal and it depends on the number and types of atoms but it is independent of the molecular structure. The two-centre $\sigma_{\text{coherent}}$ signal is dependent on the internuclear distance between two atoms. Next, we subtract the slowly varying background from the 2D-DCS maps of the ML database [33,34] and from the measured data to enhance the DCS. We note that the exact functional form for the subtraction is irrelevant as it is applied to database entries and data. For simplicity, one can use the equilibrium structure for a known molecular system. This procedure leads to clearly visible fringe patterns in the resulting difference 2D-DCS maps. We show them for the two small molecules in Fig. 1b-c. These fringe patterns are unique to the individual molecular configuration, making it significantly easier for the machine algorithm and its neural network to find the relationship between the molecular structures and their corresponding 2D-DCS maps.The database is split into three data sets to train, validate and test the model (see Fig. 1a). The training set trains the ML model to find the relationship between the molecular structures and their corresponding 2D-DCSs. The validation set is then used to assess the accuracy of the model during the training process and to determine the hyperparameters for training the model, ensuring that the trained model is not overfitting or underfitting the validation data. After the training process, the final model's quality and reliability are tested using the test set of molecules, which we have previously measured with LIED, $C_2H_2$ and $CS_2$. Once the model is validated, we use the experimental 2D-DCS map as the input for our ML model to extract the molecular structure that most likely corresponds to the measured LIED signal.

**Convolutional neural network training of ML algorithm.** Our ML algorithm utilizes a convolutional neural network (CNN) to discriminate subtle features between the 2D-DCS maps, which are used as the algorithm's input data. The architecture of CNN is composed of two parts: convolutional layers and a fully connected neural network. A 2D-DCS map is first passed through the convolutional layers to extract features using different convolution filters, convoluted across every source pixel of the input map. The filters provide various feature maps that possess distinct subtle features present in the 2D-DCS map, making the image recognition process more efficient (see Fig. 2a). The collection of feature maps is the output of the convolution process, subsequently used as the input of the fully connected neural network (see Fig. 2b). The neural network consists of layers, and each layer contains neurons (blue circles) comprising of different weighting factors, $w_i$. These weighting factors across all the connected layers in the neural network ultimately determine the relationship between the 2D-DCS and the molecular structure. The collection of features maps is first flattened to a one-dimensional (1D) array and then multiplied by each neuron's weight in the first layer. The atomic position's predicted value is calculated from all the weights in each neuron within all layers (see Fig. 2b). At the final layer, the predicted value is compared to the real value through the *cost* function (see Fig. 2c), given by

$$Cost = \frac{1}{2}(y_{pre} - y_{real})^2, \qquad (1)$$

where $y_{\text{pre}}$ and $y_{\text{real}}$ are the predicted and real values of the atom's position, respectively. This whole procedure is iterated (from Fig. 2a to Fig. 2c) to minimize the difference between the predicted and real value of the atomic position and to minimize the *cost* function. At each iteration, the filters and weights are optimized. The new optimized weight, $\omega^{\text{iter}+1}$, in each iteration is calculated by subtracting the partial derivative of the *cost* function from the current weight, $\omega^{\text{iter}}$, as shown in equation (2).

$$\omega_i^{\text{iter}+1} = \omega_i^{\text{iter}} - \alpha \frac{\partial Cost^{\text{iter}}}{\partial \omega_i^{\text{iter}}} \qquad (2)$$

Fig. 2c shows a schematic contour plot of the *cost* function with respect to the two weights ($\omega_i$ and $\omega_{i+1}$). The evolution of the *cost* value (blue dots) after five iterations is shown. The change in gradient of the *cost* function is also given by the red arrows, with the arrow length illustrating the step size taken at each iteration, called the learning rate ($\alpha$). The weights are randomly initialized at the beginning. After five iterations, we observe a decrease in the *cost* function, which indicates that the model is optimized and that the predicted values are close to the real values. In reality, our CNN model calculates a *cost* function based on thousands of parameters (e.g., weights, biases, filters) instead of only two weights, as discussed above. At each iteration, all the parameters are simultaneously updated and optimized to minimize the *cost* function and the difference between predicted and real value. This capability is unique to the ML CNN and does not exist for simple regression methods or evolutionary algorithms.

**Training and evaluation of machine learning model.** We next evaluate the accuracy of the predicted molecular structure generated by our ML model during and after the training process. Training, validation, and testing sets are generated from the normalized difference DCS maps and their molecular structures (see Supplementary sections I and II and figure S1). We use the mean absolute error (MAE) during the training process, also known as the prediction error, to evaluate the model's accuracy using the training and validation data. Fig. 3a shows the reduction in the MAE with an increasing number of times an entire dataset is iterated through the neural network during the model's training, referred to as the iteration number. With increasing iteration number, the MAE converges to a constant value of ~0.016 for both the training and validation sets of data, confirming that the model is well-trained, predicting a structure very similar to that of the input. Moreover, a similar MAE for both sets of data signifies that the model does not overfit or underfit the data. Once we find a converged MAE, we use the test data to evaluate the ML's reliability and accuracy after the training process assuming no knowledge of the input molecular structure to mimic our experimental data. Here, we use the 2D-DCSs of the test set as our input to generate the predicted structures. These predicted structures are then compared to the molecular structures from the test set of the input data. We obtain a MAE of ~0.015 using the test data (red cross in Fig. 3a), which is in good agreement with the converged MAE value (~0.016) achieved at the end of the training process using the training and validation datasets. This MAE value confirms that our final ML model is accurate and reliable. We then predict the molecular structure from our measured DCS using a normalized difference map as our ML model's input. The normalized difference map comprises the normalized experimental 2D-DCS subtracted by the normalized theoretical 2D-DCS of the equilibrium structure[33,34]. The normalization process ensures that the experimental and theoretical DCS values are on the same order of magnitude. We also avoid the need for a fitting factor as is typically used in QRS-LIED (see Supplementary section III). The molecular structure predicted by our ML model and the experimental DCS input is then used to calculate its corresponding theoretical 2D-DCS. We then evaluate the correlation between the normalized theoretical and experimental 2D-DCS using the Pearson correlation (see Supplementary section IV). We obtain a Pearson correlation value of 0.94 (see Fig. 3b), which indicates that both DCSs are strongly correlated and that the predicted molecular structure is accurate and reliable.

**Extracting measured molecular structure with machine learning.** As a first step, we test the ML framework by applying it to the small linear symmetric top and non-linear asymmetric top molecules, whose structure we have previously determined with LIED. We extract the molecular structure of $CS_2$ and $C_2H_2$ from the experimental 2D-DCS using our ML model trained on five separate datasets. Figure 3c shows the average structural parameters of the so determined structure for $C_2H_2$ ($CS_2$) of $R_{CC} = 1.23 \pm 0.11$ Å and $R_{CH} = 1.08 \pm 0.03$ Å ($R_{CS} = 1.87 \pm 0.14$ Å and $\theta_{SCS} = 104.7 \pm 6.4°$). These values are in excellent agreement with the values retrieved by the QRS model; see the comparison in Tab. 1. After training our ML model on five separate datasets, the predicted structures vary slightly. This is because the neural network during the training process uses a random number generator to select the input-target pairs of data. Thus, each neuron's corresponding initial weights and biases are also randomly chosen, leading to slightly different starting conditions in the CNN training of the ML model. Using random initial conditions for the neurons ensures that systematic errors are minimized. In addition, we generated predicted structures from the ML model, trained on five separate occasions, to ensure the reliability of the predicted values. We find that the uncertainty in our ML-predicted structural parameters arise from two contributions: the predicted model error and the experimental statistical error. The predicted model error is obtained by calculating the MAE between the absolute and calculated value of the structural parameter using the test set as described earlier. The experimental error arises from the standard error in our experimental DCS, following a Poissonian statistical distribution (see Supplementary section V). We include the experimental error into the predicted values using a modified experimental DCS that contains the extrema of the experimental error. Then, we study the variance in the predicted structure that includes the experimental error relative to the original unmodified experimental DCS. This procedure leads to extracted structures with picometre accuracy.

**Extracting complex 3D molecular structure.** With the successful retrieval of the smaller 1D and 2D molecules, we put our ML framework to the test to extract the structure of gas-phase (+)-Fenchone ($C_{10}H_{16}O$; 27 atoms), measured with LIED. For such a large and complex 3D molecule, the ML has the decisive advantage to interpolate and learn between the course grid of pre-calculated structures and to take into account a manifold of degrees of freedom in the solution space. Thus, we can use a sufficiently reduced database that only considers (i) four groups of atoms of the molecule (see inset of Fig. 4A) and (ii) a molecule-wide global change in structure. Next, we train our ML model to find the relationship between the molecular structures and corresponding 2D-DCSs with the such reduced database. This approach drastically minimizes computational time. Figure 4A shows the MAE achieved at each iteration number using the neural network, convolved with training and validation sets of simulated data. This achieves a MAE of 0.02. Consequently, Fig. 4B shows a strong correlation between the experimental and predicted theoretical 2D-DCS with a Pearson correlation coefficient of 0.94. As example, Fig. 4C shows the extracted $(x, y, z)$ 3D cartesian coordinates of seven atoms of (+)-Fenchone (green circles). We find that the ML-LIED-measured (+)-Fenchone structure shows only slight deviations from the equilibrium ground-state neutral molecular structure (red triangles) which are involuntarily caused by the presence of the LIED laser field. The degree of uncertainty of the predicted 3D positions (green circles) are shown on top of the predicted 3D molecular structure in Fig. 4D. This shows that ML-LIED is capable to extract a complex 3D molecular structure such as (+)-Fenchone.

**Discussion**

This work establishes a ML-based framework to overcome present limitations in structural retrieval from diffraction measurements. The problem with present methods is the need to compare a measured diffraction pattern with a pre-calculated structure and the extremely poor scaling of pattern matching methods with the quickly increasing number of degrees of freedom of larger complex molecular structures. This is compounded by the need pre-calculate a very large set of molecular configurations in different orientations and with high resolution. Further, identification of such a pre-calculated set reduces to finding a global extremum in a multi-dimensional solution space which is a difficult inverse problem to solve. These issues are tractable for small molecular systems and for systems where dimensionality of the problem can be reduced significantly. This is however not possible for large and complex molecular structures and the total calculation time scales as $n \times 3^N$, where $N$ is the number of atoms and $n$ is the number of steps. To put this into perspective, it takes approximately five minutes to calculate a single 2D-DCS map for (+)-Fenchone on a standard desktop computer (i3 Intel processor, 8 GB RAM). A 20-atom system with $n = 5$ steps will require an unrealistic $1.4 \times 10^9$ hours of calculation time. Thus, it is not feasible at the present time to extract complex molecular structure via calculating all possible configurations for all degrees of freedom. To overcome the unfavourable scaling of the problem, we make use of the fact that a ML framework can simultaneously identify a large multitude of features by pattern matching on an interpolated dataset, despite having initially trained the ML framework on a coarse ensemble of structures. We showed that this permits ML-LIED to reveal the 3D location of each atom in the molecule, providing significantly more detailed structural information than any other method that relies on the identification of non-overlapping peaks in the scattering radial distribution function. Such identification is near impossible for larger and complex molecular structures due to the large number of peaks in the radial distribution that overlap due to the multitude of unresolvable two-atom combinations. The reduced computational demands for complex molecules may prove decisive to address transient structures and transition states of complex molecules. ML-LIED has the advantage that once the ML model is validated, the molecular structures can be identified for each experimental 2D-DCS map, avoiding the high computational cost of repeated chi-square fittings as in time-resolved QRS-LIED. Lastly, using 2D-DCSs as an input instead of 1D-DCSs utilizes the complete measured molecular interference signal thus maximizing confidence in the identification of the measured molecular structure. We also note that the ML framework draws from unambiguous pattern matching conditions. It is conceivable, though very unlikely, that different non-equilibrium structures may be identified as the same structure. The ML framework is amendable to be adapted and trained on the existence of a combined total interference signal from two or more molecular structures. This may allow to predict the multiple molecular structures contributing to the total measured interference signal. It should be noted that the same problem of contributions from multiple molecular structures to the measured signal can also arise in other LIED and UED methods, and it is not unique to just ML-LIED. This possibility for ML-LIED may overcome standing problems in time-resolved UED studies of isomerization or ring-opening reactions, e.g., where the molecular structure changes will lead to subtle changes in the measured 2D-DCS. Thus, the combination of LIED with a machine learning framework and CNN provides a powerful new opportunity to determine the structure of large molecules.

**Methods**

**LIED data.** In this work, we used experimental data for $C_2H_2$[16], $CS_2$[20] and (+)-Fenchone which were measured with a reaction microscope[35] to demonstrate our ML algorithm. We calculated the 2D-DCS of 40,000 - 120,000 possible molecular structures in our database using the IAM.

**Machine learning framework.** We generate thousands of structures and calculated their 2D DCS as an input dataset. Our convolutional neural network contained 3 convolutional layers and 30 fully connected layers, with the first convolutional layer containing 32 filters with kernel size 5 x 5, the second layer has 32 filters with kernel size 3 x 3, the third has 32 filters with kernel size 3 x 3. A batch size of 120 was used, and batch normalization is used to avoid overfitting. Our model was trained and validated with an iteration number of greater than 50 to generate a model that does not overfit or underfit the input data. The training of the ML model costs one-two hours to run on the Google cloud GPU (NVIDIA® Tesla® V100). To predict a molecular structure with an experimental 2D-DCS input requires less than one minute of calculation time.

**Data availability**

The data that support the findings of this study are available from the corresponding author upon reasonable request.

**Code availability**

The codes used in this study are available from the corresponding author upon reasonable request.

| | Table 1 Machine learning $C_2H_2$ and $CS_2$ predicted structures | | | |
|---|---|---|---|---|
| | Parameter | Equilibrium | QRS | ML |
| $C_2H_2$ | $R_{CC}$ (Å) | $1.20^{34}$ | $1.24 \pm 0.04^{16}$ | $1.23 \pm 0.11$ |
| | $R_{CH}$ (Å) | $1.06^{34}$ | $1.10 \pm 0.03^{16}$ | $1.08 \pm 0.03$ |
| $CS_2$ | $R_{CS}$ (Å) | $1.55^{33}$ | $1.86 \pm 0.23^{20}$ | $1.87 \pm 0.14$ |
| | $\theta_{SCS}$ (Å) | $180^{33}$ | $104 \pm 20.2^{20}$ | $104.7 \pm 6.4$ |
| The machine learning (ML) predicted results are compared to those retrieved by the quantitative rescattering (QRS) model[16,20] and to the equilibrium structure[30,31] of the corresponding neutral molecule in its ground electronic state. | | | | |

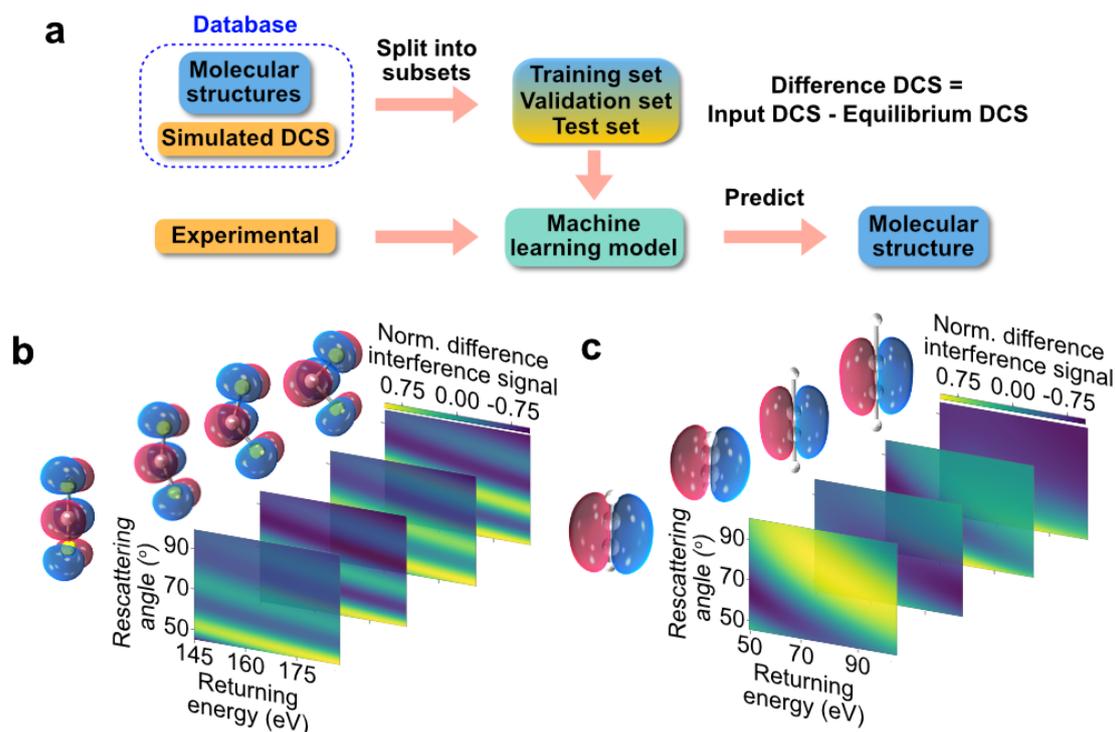

**Fig. 1 Machine learning schematic. a** A database of molecular structures and their corresponding simulated 2D-DCSs are split into three sets: training, validation, and test sets. Here, the simulated 2D-DCS maps are calculated *via* the independent atomic model (IAM). Once the machine learning (ML) model is validated, the experimental 2D-DCS map is used as an input to predict the molecular structure that most likely contributes to the measured interference signal. **b,c** Exemplary simulated difference 2D-DCS maps for four structures of carbon disulfide (b) and acetylene (c) calculated from subtracting the corresponding 2D-DCS of the equilibrium molecular structure.[33,34] Fringe patterns are visible, enhancing the difference in the elastic scattering signal for the different molecular structures. The molecular coordinates (used as labels in ML algorithm) for $C_2H_2$ (a) and $CS_2$ (b) molecules in cartesian and polar coordinates, respectively.

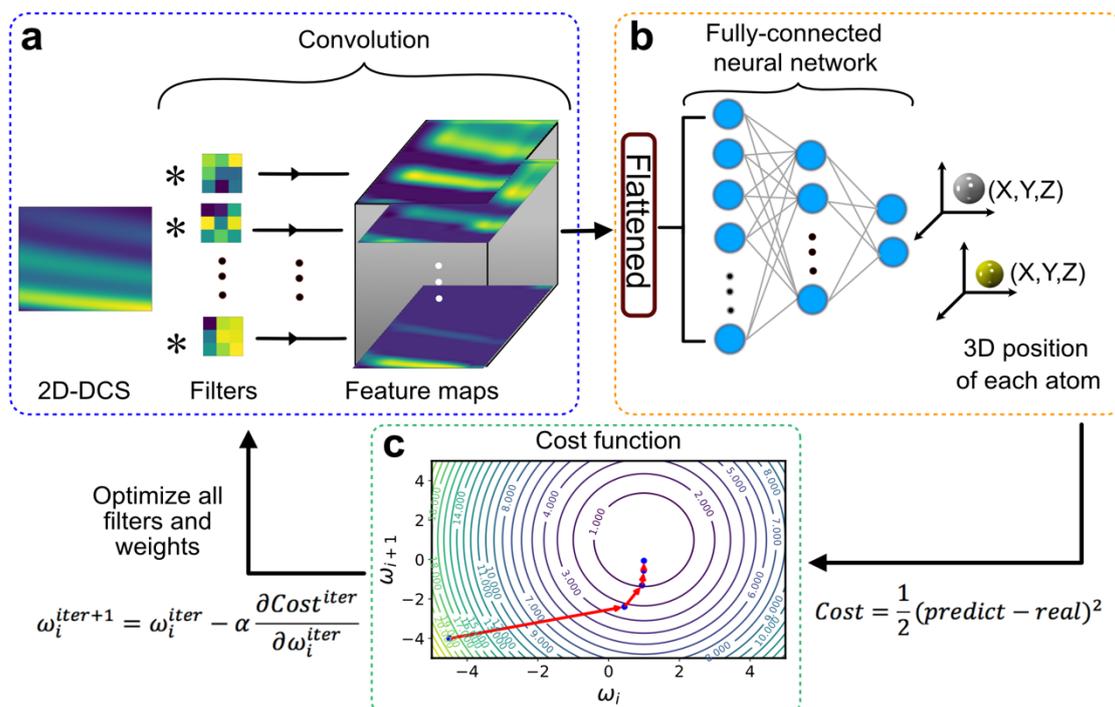

**Fig. 2 CNN training of ML algorithm to predict molecular structure. (a)** The 2D-DCS is convoluted by different filters to generate a collection of feature maps. **(b)** The collection of feature maps is first flattened into a one-dimensional array and multiplied by the weights in each neuron of all layers to predict the atomic position for each atom in the molecule. **(c)** Contour plot of the *cost* function for two weights ($\omega_i$ and $\omega_{i+1}$). The blue dot represents the *cost* function value, and the red arrows show the direction of the gradient of the *cost* function. After five iterations, the *cost* function is minimized.

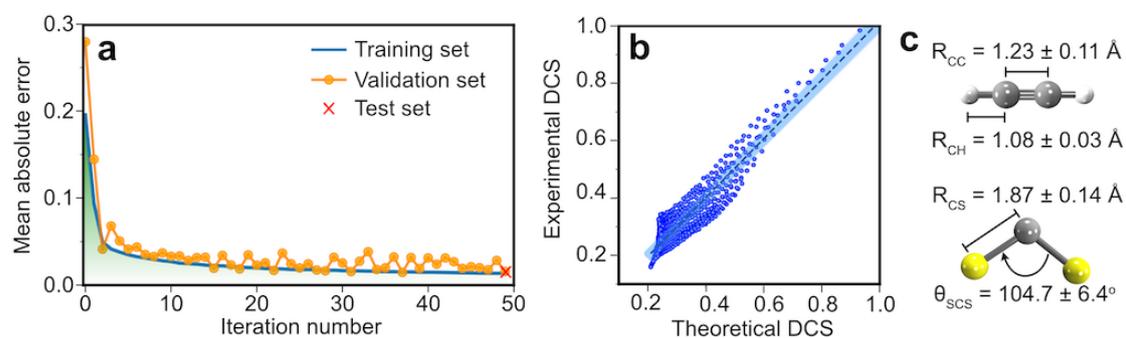

**Fig. 3 Evaluating machine learning results during and after the training process. a** Mean absolute error (MAE) obtained for each iteration number that the neural network convolutes the training and validation sets of our simulated data. The MAE obtained with the test data is indicated by the red cross. **b** Correlation between normalized experimental and theoretical two-dimensional differential cross-section (2D-DCS), reshaped into a 1D array. A linear fit (blue dashed line) and the confidence interval (blue shaded area) using the bootstrapping method are applied to the data. A Pearson correlation coefficient of 0.94 is obtained, indicating a strong correlation between the two data. **c** Predicted structural parameters obtained by machine learning.

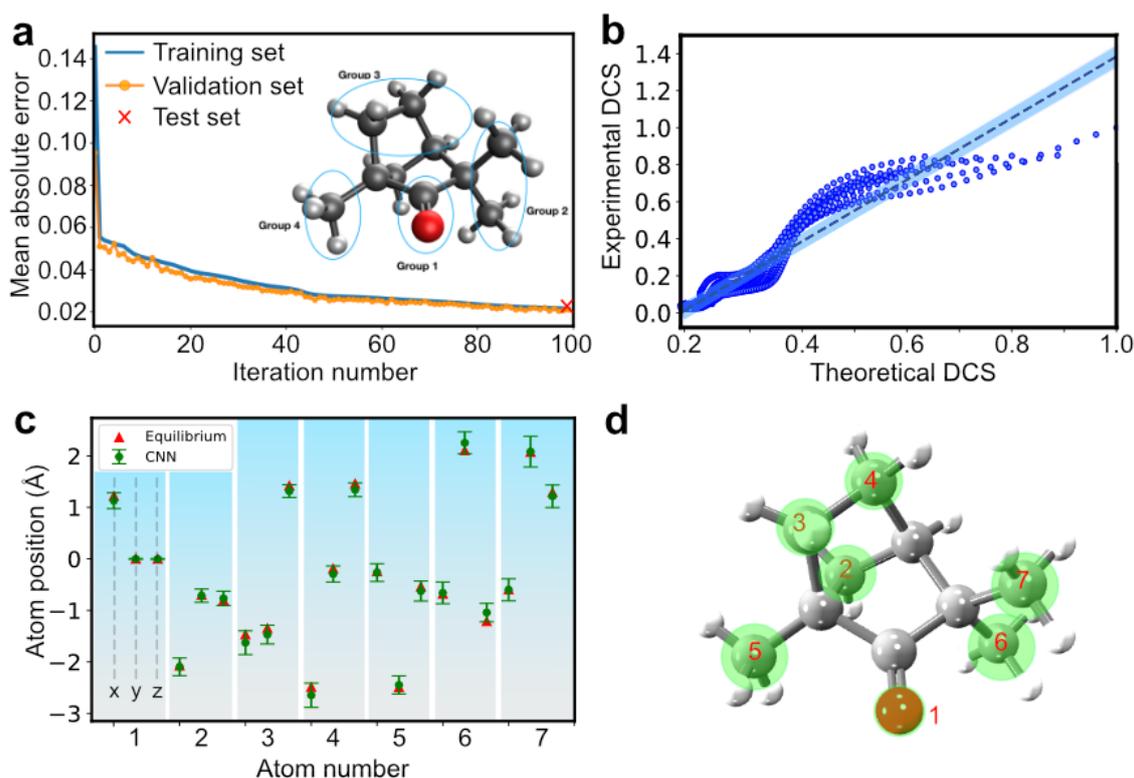

**Figure 4 | Extracted (+)-Fenchone molecular structure by machine learning. (a)** Mean absolute error (MAE) at each iteration number using a neural network convoluted with the training and validation sets of simulated data. The MAE obtained with the test data is indicated by the red cross. The inset shows a schematic of the four groups of the (+)-Fenchone molecule which the ML algorithm has been trained on. **(b)** Correlation between normalized experimental and theoretical 2D-DCS similar to Fig. 3B. A linear fit (dashed lined) and the confidence interval (blued shaded area) using the bootstrapping method are applied to the data. A Pearson correlation coefficient of 0.94 is obtained, indicating that a strong correlation between the two data exists. **(c)** The predicted $(x, y, z)$ 3D positions for seven atoms in (+)-Fenchone using ML model (green circles). The equilibrium ground state 3D positions of neutral (+)-Fenchone are shown (red triangles). **(d)** Schematic of the predicted 3D molecular structure. The green circles indicate the area of uncertainty.

## Acknowledgements

J.B. and group acknowledge financial support from the European Research Council for ERC Advanced Grant "TRANSFORMER" (788218), ERC Proof of Concept Grant "miniX" (840010), FET-OPEN "PETACom" (829153), FET-OPEN "OPTOlogic" (899794), Laserlab-Europe (EU-H2020 871124), MCIN for PID2020-112664GB-I00 (AttoQM); AGAUR for 2017 SGR 1639, MINECO for "Severo Ochoa" (SEV- 2015-0522), Fundació Cellex Barcelona, the CERCA Programme / Generalitat de Catalunya, and the Alexander von Humboldt Foundation for the Friedrich Wilhelm Bessel Prize. We also acknowledge Marie Sklodowska-Curie Grant Agreement 641272. X.L. and J.B. acknowledge additional financial support from China Scholarship Council.

**Author contributions**

J.B. conceived and supervised the project. X. L. analyzed the data and developed the machine learning algorithm. X.L. and A.S. generated the input database. X.L., K.A., A.S., B.B., T.S. and J.B. contributed to the data interpretation and writing of the manuscript.

**Competing interests**

The authors declare no competing interests.